\documentclass[12pt,letterpaper]{article}
\usepackage[letterpaper,ignoreall,top=2.65cm,bottom=2.7cm,left=2.65cm,right=2.65cm,foot=1cm,head=1.5cm]{geometry}
\usepackage{setspace}
\usepackage{xcolor}

\usepackage{times}
\usepackage{amsmath}
\usepackage{amssymb}
\usepackage{bm}
\usepackage{here}
\usepackage{graphicx}
\usepackage{framed}
\definecolor{shadecolor}{rgb}{0.9,0.9,0.9}
\usepackage[utf8x]{inputenc}

\usepackage{rsc}
\setcitestyle{square}

\usepackage[below]{placeins}

\newlength{\figurewidth}
\begin{document}  

\date{\today}

\begin{center}
\Large\textbf{Rational Design of Superconducting Metal Hydrides via Chemical Pressure Tuning}
\end{center}
\begin{center}
\textit{Katerina P. Hilleke and Prof. Eva Zurek$^*$} \\[1ex]
Department of Chemistry, State University of New York at Buffalo, \\
Buffalo, NY 14260-3000, USA \\[1ex]
E-mail: ezurek@buffalo.edu\\
Twitter: @UBChemistry \\
\end{center}

\section*{Abstract}
The high critical superconducting temperatures ($T_c$s) of metal hydride phases with clathrate-like hydrogen networks have generated great interest. Herein, we employ the Density Functional Theory-Chemical Pressure (DFT-CP) method to explain why certain electropositive elements adopt these structure types, whereas others distort the hydrogenic lattice, thereby decreasing the $T_c$. The progressive opening of the H$_{24}$ polyhedra in MH$_{6}$ phases is shown to arise from internal pressures exerted by large metal atoms, some of which may favor an even higher hydrogen content that loosens the metal atom coordination environments. The stability of the LaH$_{10}$ and LaBH$_{8}$ phases is tied to stuffing of their shared hydrogen network with either additional hydrogen or boron atoms. The predictive capabilities of DFT-CP are finally applied to the Y-X-H system to identify possible ternary additions yielding a superconducting phase stable to low pressures. 
\newpage

\section*{Introduction}
Many diamonds have been broken and CPU hours have been burned in the search for a conventional room temperature superconductor \cite{Zurek:2021k}. Currently, the most promising pathway towards such a material utilizes \emph{chemical precompression} \cite{Ashcroft:2004a}, where doping hydrogen by a second element applies an internal pressure that stabilizes the exotic metallic phase at much lower pressures than in solid hydrogen. This strategy has borne fruit, with high superconducting critical temperatures, $T_c$s, reported for H$_3$S ($T_c$ = 203~K, 150~GPa) \cite{Drozdov:2015a}, LaH$_{10}$ ($T_c$ = 260~K, 200~GPa) \cite{Somayazulu:2019,Drozdov:2019}, YH$_9$ ($T_c$ = 262~K, 182~GPa) \cite{Zurek:2020j}, YH$_6$ ($T_c$ = 224~K, 166~GPa) \cite{Troyan:2021a}, and in carbonaceous sulfur hydride ($T_c$ = 288~K, 270~GPa) \cite{Snider:2020a} -- all detected at pressures lower than those required to metallize pure hydrogen \cite{Monacelli:2021}. Increasing chemical complexity may decrease these pressures even further, hinted at by the prediction of superconducting LaBH$_{8}$ ($T_c$ = 126~K at 50~GPa) \cite{DiCataldo:2021} and LaBeH$_{8}$ ($T_c$ = 191~K at 50~GPa) \cite{Zhang:2021} phases.

\begin{figure*}
\begin{center}
\includegraphics[width=1.2\figurewidth]{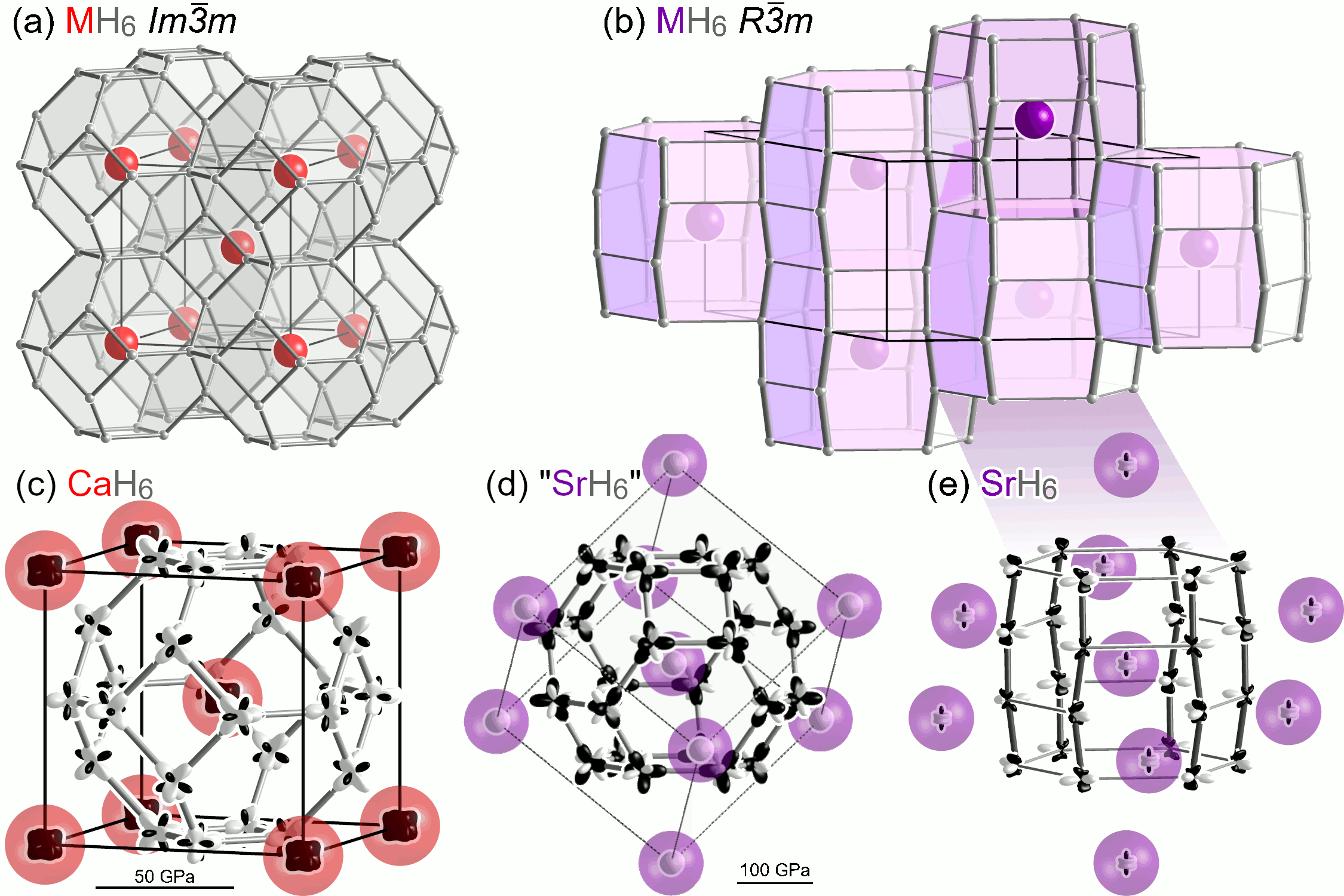}
\end{center}
\caption{The predicted (a) $Im\bar{3}m$ and (b) $R\bar{3}m$ symmetry metal hexahydrides. Chemical pressure schemes for (c) $Im\bar{3}m$ CaH$_{6}$, (d) $Im\bar{3}m$ SrH$_{6}$, and (e) $R\bar{3}m$ SrH$_{6}$. Negative pressure, or a local desire for contraction of the coordination environment, is represented by black lobes, and positive pressure, indicating a local desire for expansion of the coordination environment, is represented with white lobes. The magnitude of the pressure in a particular direction is indicated by the distance from the center of the atom to the surface of the lobe along that direction. 
\label{fig:MH6}}
\end{figure*}

Two families of structures account for most high-$T_c$ compounds identified to date: those containing covalent bonds between hydrogen and a main group element~\cite{Yao:2018}, and those with weakly covalently bonded hydrogenic frameworks reminiscent of clathrate cages stuffed with electropositive
elements \cite{Peng:2017a,Zurek:2018m,Zurek:2016j}. Our focus is on the latter, with a prominent example being the $Im\bar{3}m$ MH$_6$
hexahydrides, built from a bcc packing of face-sharing M@H$_{24}$ polyhedra that each contain six square and eight hexagonal faces
(Figure~\ref{fig:MH6}a). Many compounds have been predicted to adopt this structure under pressure, including those with M=Ca, Mg, and
Sc.~\cite{Wang:2012, Feng:2015a, Peng:2017a, Abe:2017, Zurek:2018b, Qian:2017} Two more ubiquitous structure types -- $P6_3/mmc$ MH$_9$ and $Fm\bar{3}m$ MH$_{10}$ -- are exemplified by YH$_9$~\cite{Li:2019} built from Y@H$_{29}$ clusters, and LaH$_{10}$~\cite{Liu:2017, Peng:2017a} composed of La@H$_{32}$ clusters in an fcc network.

In addition to these highly symmetric archetypes, several structures based on their distortions have been predicted.  MH$_6$ compounds that contain larger M atoms (such as Sr \cite{Hooper:2013,Wang:2015} and La \cite{Peng:2017a,Liu:2017}) favor $R\bar{3}m$ symmetry with M@H$_{24}$ polyhedra in which opposite hexagonal faces are widened, resulting in channels wherein helical chains of H atoms comprise the walls (Figure~\ref{fig:MH6}(b)). Even more drastic distortions are predicted to occur in the preferred BaH$_6$ phase with $Imm2$ symmetry.~\cite{Hooper:2012} 
$C2/m$ CaH$_{9}$ is a distortion of $P6_{3}/mmc$ MH$_9$ \cite{Shao:2019}. Finally, complex phase behavior has been explored in MH$_{10}$ compounds, with $R\bar{3}m$ and $C2/m$ structures \cite{Shao:2019,Liu:2018,Semenok:2018,Wang:2015,Geballe:2018}  -- all distortions of the highly symmetric $Fm\bar{3}m$ phase -- predicted to be stable in different pressure regimes within the static lattice approximation. Calculations including quantum and anharmonic effects, on the other hand, smooth the potential energy surface for  LaH$_{10}$ so that $Fm\bar{3}m$ becomes the ground state~\cite{Errea:2020a}.

A major factor in the structural preferences among the metal clathrate hydrides -- hinted at above -- is the importance of a good match between the sizes of the metal atom and the hydrogenic cluster encapsulating it. Each of the aforementioned distorted geometries can be empirically related to a metal atom that is too large or too small for the coordination environments in the $Im\bar{3}m$, $P6_{3}/mmc$, and $Fm\bar{3}m$ ``parent'' crystal lattices. Because these distortions lower the density of states at the Fermi level, thereby decreasing the $T_c$, it is important to understand their origin. The Density Functional Theory-Chemical Pressure (DFT-CP) method~\cite{Fredrickson:2012}, which has been applied successfully to explain the stability and structural distortions in a plethora of complex inorganic materials \cite{Hilleke:2017,Lobato:2019,Zurek:2020e}, visualizes the internal stresses inherent in crystal structures as a consequence of steric constraints. It is therefore the ideal technique to analyze how chemical substitution affects the structures the superconducting metal hydrides adopt.

Herein, we employ the DFT-CP method to uncover the relationship between the size of the metal atom and the structure of the hydrogenic lattice, focusing on $Im\bar{3}m$ versus $R\bar{3}m$ MH$_6$. Moreover, we show that the DFT-CP technique can identify the most promising open sites in known superhydrides, and elements that could be incorporated within them~\cite{Lobato:2019}, so the hydrides can be quenched to lower pressures while at the same time retaining their propensity for superconductivity. One structure we pinpoint in such a way, YC$_2$H$_8$, is metastable at 50~GPa where its $T_c$ is estimated to be $\sim$60~K. Our study demonstrates how to rationally design warm and light superconductors derived from known binaries without the need of employing first-principles based crystal structure prediction searches, prohibitively expensive for multicomponent systems because of their combinatorial complexity.

\section*{Results and Discussion}
To get a better idea of the internal tensions existing in a compound that is thermodynamically and dynamically stable under pressure, we focused on $Im\bar{3}m$ CaH$_{6}$, the first metal hexahydride to be predicted \cite{Wang:2012}. Figure~\ref{fig:MH6}(c) presents its integrated CP scheme, decorated with atom-centered spherical harmonic functions that represent the magnitude and sign of the chemical pressure experienced by each atom in various directions. The Ca atoms (red) possess negative pressure features, which are slightly elongated along the nearest neighbor Ca-Ca contacts, in all directions. This indicates that a tighter coordination environment around Ca would be preferred. In addition, negative pressure lobes appear on the hydrogen atoms, reflecting a desire for shortening the Ca-H distances. These preferences are stymied, however, when we evaluate the situation within the hydrogen sublattice. Along the H-H contacts, positive pressure features show that the hydrogen lattice is already too contracted. Further tightening the coordination environment around the Ca atoms and shortening the Ca-H contacts would force the H atoms into even closer proximity with one another, exacerbating the already too-small hydrogen lattice. Thus, negative Ca-H pressures are balanced against positive H-H pressures, with Ca overall being slightly small for its coordination environment within the H$_{24}$ cage.

Based on the tensions present in the CP scheme of CaH$_{6}$, we can evaluate the internal stresses that occur when Ca, with a non-bonded atomic radius of 2.31~\AA, is replaced by a much larger atom such as Sr (2.49~\AA). The CP scheme of the $Im\bar{3}m$ parent of SrH$_{6}$ (Figure~\ref{fig:MH6}(d)), shows a drastically different situation than found in CaH$_{6}$. In contrast to the negative pressures decorating the Ca atoms, the Sr atoms display a near isotropic positive pressure, indicating that Sr is too large to comfortably fit the H$_{24}$ polyhedral cage. On the H sublattice, positive pressure features are directed towards the Sr atoms, reinforcing the too-tight fit of Sr in the cage. This constrained environment is maintained due to the presence of negative pressures along the H-H contacts comprising the edges of the cage, which show an H lattice that is overly stretched in an attempt to fit the large Sr atom. The CP schemes of  $Im\bar{3}m$ parent models of BaH$_{6}$ and LaH$_{6}$  paint a very similar picture: positive pressures on the Ba (2.68~\AA{}) and La (2.43~\AA) atoms and along Ba-H and La-H contacts are balanced against negative H-H pressures in the clathrate cages (Figure S1). For Sr-, Ba-, and LaH$_{6}$, the resulting CP schemes are essentially the reverse of the CP scheme of CaH$_{6}$ -- too large versus slightly small metal atoms and an overly stretched, instead of a compressed hydrogenic lattice.  Thus, Sr, Ba, and La would benefit from a looser coordination environment than the one provided by the [4$^{6}$6$^{8}$] H$_{24}$ polyhedra with $Im\bar{3}m$ symmetry. 

The $R\bar{3}m$ structure favored by SrH$_{6}$ relieves the steric tensions present within $Im\bar{3}m$, but how does it do so? The former can be derived from the latter by elongating four out of the six shortest metal-metal contacts and slightly deforming the face that bisects them so it is no longer an ideal hexagon \cite{Hooper:2013}. The resulting distorted H$_{24}$ polyhedra no longer maintain equal distances between the H atoms, instead condensing into helical chains running along the \emph{c} axis. As illustrated in Figure~\ref{fig:MH6}(b), this structural distortion reflects an opening of the  [4$^{6}$6$^{8}$] H$_{24}$ polyhedra, which can still be identified, although with larger distances between hydrogens in neighboring helices than along the helices. 

These structural changes lead to a very different CP scheme for $R\bar{3}m$ SrH$_{6}$ than for the $Im\bar{3}m$ parent. In the distorted structure the Sr atoms no longer experience positive CP in each direction. Instead, they possess positive pressures oriented towards the nearby H atoms in the walls of the channel and negative pressures along the \emph{c} axis, which is directed along the shortest Sr-Sr contact. The hydrogen network similarly experiences CP relief upon opening of the H$_{24}$ polyhedra,  as seen in the smaller CP lobes decorating the H atoms. In condensing into helical chains the H-H distances shorten (1.24 \AA), alleviating the overly stretched contacts in the parent $Im\bar{3}m$ phase (1.51~\AA{}). The distances between H atoms in neighboring helical chains have increased (2.02 \AA), but at such a large distance their interaction is negligible and CP lobes are no longer observed along these contacts. Thus, the $Im\bar{3}m\rightarrow R\bar{3}m$ distortion allows the simultaneous contraction of the too-stretched H network and the opening of the cramped environment surrounding the Sr atoms, seen in the CP relief on all atoms. In line with its similar CP scheme, some calculations have suggested that LaH$_{6}$ also adopts $R\bar{3}m$ symmetry~\cite{Liu:2017,Peng:2017a}. One proposed BaH$_{6}$ phase with $Imm2$ symmetry~\cite{Hooper:2012} goes even further in fragmenting the hydrogenic network, leaving H$_{3}^{-}$ subunits upon greatly enlarging the Ba coordination environment. Deformations of simple structures, such as the ones described above, have been widely explored as avenues of CP relief~\cite{Berns:2014,Engelkemier:2013,Hilleke:2018}.  

These distortions significantly affect the $T_c$ of metal polyhydrides. Calculations predict that the superconductivity in phases with 3D hydrogenic clathrate lattices persists to higher temperatures than in phases with lower-dimensional H networks or with molecular motifs~\cite{Zurek:2018m} (\emph{cf.}\ 235~K at 150~GPa for CaH$_6$~\cite{Wang:2012} versus 156~K at 250~GPa for SrH$_6$~\cite{Bi:2019a}).

\begin{figure*}
\begin{center}
\includegraphics[width=1.3\figurewidth]{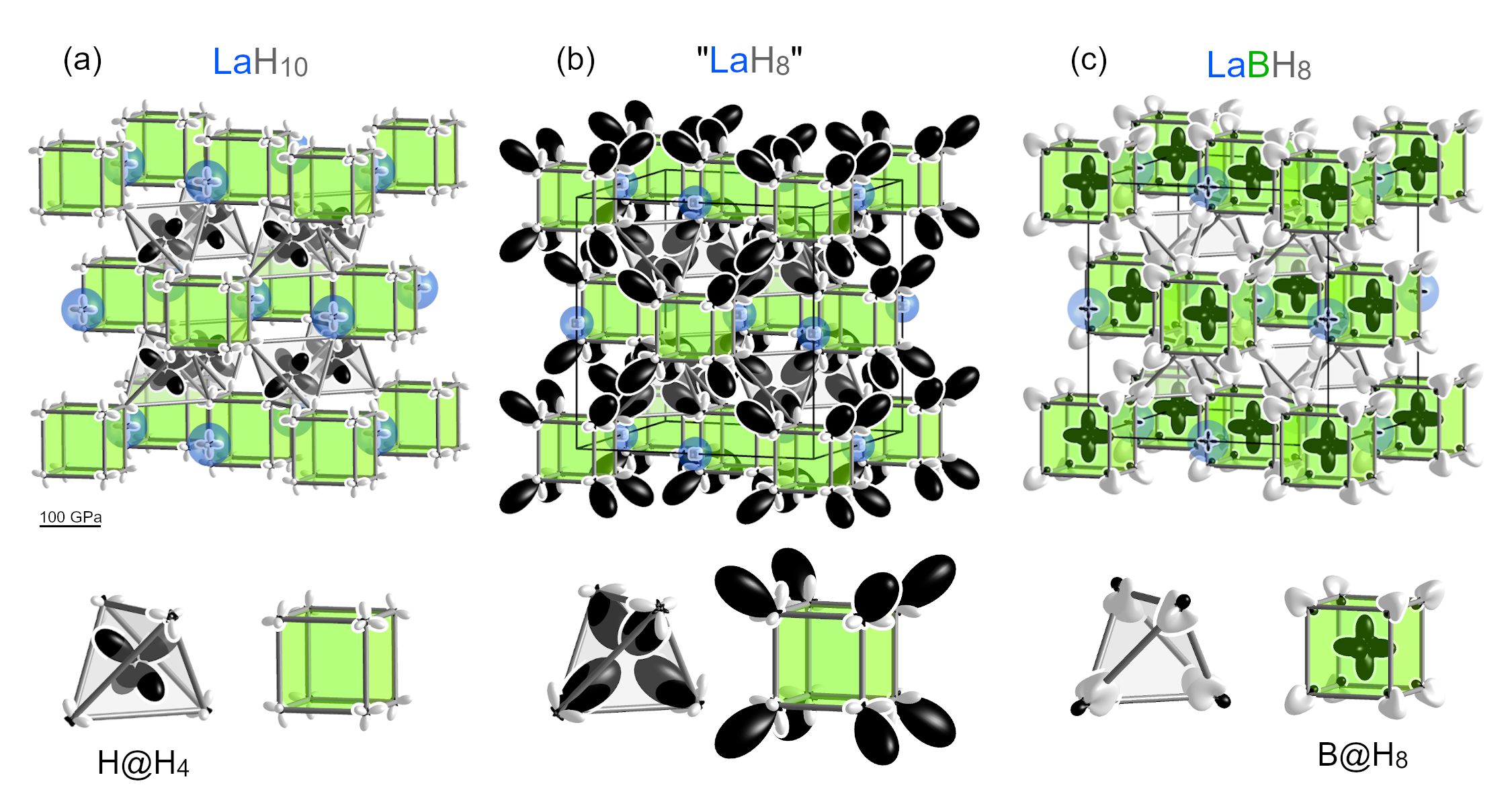}
\end{center}
\caption{Chemical pressure schemes of lanthanum hydrides with $Fm\bar{3}m$ lattices of vertex-sharing H$_{4}$ tetrahedra and H$_{8}$ cubes (insets). In LaH$_{10}$ (a), the tetrahedra are stuffed with additional H atoms, leaving some positive H-H pressures within the tetrahedra but optimized pressures along the cube edges. Removing the additional H atoms yields the model ``LaH$_{8}$" phase (b), where large negative pressures inside the empty tetrahedra indicate the need for additional atoms to fill out the hydrogenic network that is stretched to accommodate the large La atoms. In LaBH$_{8}$ (c), the H$_{8}$ cubes are stuffed with additional boron atoms that display negative pressure, compressing the tetrahedra (seen in the positive pressures within) but achieving very small residual pressures on the La atoms.
\label{fig:LaH10_LaBH8_CP}}
\end{figure*}

A third and final way to relieve the CP in a too-tight or too-loose coordination environment is by adding or removing atoms to shrink or expand it as needed. This rationale can be used to explain why $Fm\bar{3}m$ LaH$_{10}$ is stable and has been synthesized, whereas the hypothetical $Im\bar{3}m$ LaH$_6$ is not. The coordination environment within $Fm\bar{3}m$ LaH$_{10}$ is constructed of a H$_{32}$ polyhedron with six square and twelve hexagonal faces, [4$^6$6$^{12}$]. Another way to view the LaH$_{10}$ lattice is as  a network of vertex-sharing H$_{4}$ tetrahedra and H$_{8}$ cubes (Figure~\ref{fig:LaH10_LaBH8_CP}a) that adopt a fluorite-like packing arrangement with the tetrahedra on the anionic and the cubes on the cationic sites. The La atoms lie within the octahedral vacancies in the H$_{8}$ cubes. The addition of hydrogen to the network allows for shorter H-H distances in LaH$_{10}$ (1.34 and 1.48~\AA{}) as compared to the $Im\bar{3}m$ LaH$_{6}$ parent (1.58~\AA), ameliorating the steric strain of enclosing the large La atom. Accordingly, the CP scheme (Figure~\ref{fig:LaH10_LaBH8_CP}a) illustrates that the H-H contacts in LaH$_{10}$ are largely satisfied. The most prominent feature in the CP scheme are lobes pointing from the H to the La atoms. The 24 nearest neighbor hydrogen (H1) atoms display smaller positive pressures and the eight second nearest neighbor hydrogens (H2) large negative pressures, hinting that an ideal La-H distance may be somewhere in between. Along the edges of the cubes and tetrahedra (Figure~\ref{fig:LaH10_LaBH8_CP}a, inset), minimal CP features reflect fairly optimized distances within the H1 lattice, with only slight positive pressures between H1 atoms comprising the tetrahedra and H2 atoms stuffing them. The H-H distances along the cube edges are 1.48 \AA, lengthening to 2.19 \AA~along the tetrahedron edges to accommodate the stuffing atom.

The role of the stuffed hydrogen becomes clear upon examination of a model ``LaH$_{8}$" structure that is constructed by removing this atom from the LaH$_{10}$ lattice. The most striking features in this hypothetical structure (Figure~\ref{fig:LaH10_LaBH8_CP}b) are the large negative pressure lobes directed towards the empty center of the H$_{4}$ tetrahedra. As in LaH$_{10}$, the CPs along the H-H contacts forming the cube and tetrahedron edges are relatively small and satisfied, but now the lack of a centering atom within the tetrahedra has left a void within the structure that is highlighted by the strong negative CP features. In this model, the La atoms display positive pressure, indicating that the hydrogenic lattice contracts to relieve the void space left by the lack of stuffing atoms, leaving a too tight coordination environment for La. Although a phase with LaH$_{8}$ stoichiometry has been predicted at 150 GPa~\cite{Liu:2017}, it has a relatively low $C2/m$ symmetry and a La coordination environment that mixes the clathrate-like cages of $Im\bar{3}m$ LaH$_{6}$ and $Fm\bar{3}m$ LaH$_{10}$.  Thus, the presence of the additional H atoms in the LaH$_{10}$ network is needed to stabilize the expanded scaffolding around the La atom. 

Rather than stuffing the H$_{4}$ tetrahedra in ``LaH$_8$" with hydrogen, another way to stabilize the hydrogenic framework is to expand the H$_{8}$ cubes, thereby compressing the tetrahedra they share vertices with -- as exemplified by the recently proposed ternary LaBH$_8$ phase (Figure~\ref{fig:LaH10_LaBH8_CP}c).~\cite{DiCataldo:2021} In this phase a boron atom is placed inside each H$_{8}$ cube and the tetrahedra are empty, resulting in an expansion (shrinking) of the H1-H1 contact along the cube (tetrahedral) edges to 1.71~\AA~(1.86~\AA{}). In fact, this leaves positive pressures along the edges of the tetrahedra, showing that expansion of the H$_{8}$ cubes upon boron incorporation has slightly overcompressed the void space within them. The CP features on the La atoms are very small, indicating that the coordination environment supported by boron inclusion is of appropriate size. The most prominent features in this CP map, therefore, are the negative lobes decorating the B atoms. Negative CP has previously been correlated with soft atomic motions~\cite{Engelkemier:2016}, with cases where an atom bears negative CP in all directions, such as those seen here, as a possible indicator of a rattling atom~\cite{Guo:2016}. In fact, it has been pointed out that such a dispersionless rattling mode of the boron atoms within their H$_8$ cubes is evident in the phonon band structure of  LaBH$_{8}$, as described in Ref.~\cite{DiCataldo:2021}.

Thus, the hydrogenic network of vertex-sharing cubes and tetrahedra can be stabilized to incorporate La atoms by stuffing either set of holes. The placement of boron atoms within the cubes or hydrogen atoms within the tetrahedra achieves a successful scaffolding, but other elements of different sizes may be more appropriate for other systems. On its own, the cube-and-tetrahedron network is too small to contain La atoms without becoming overly stretched -- as seen in the CP scheme of the ``LaH$_{8}$" model in Figure~\ref{fig:LaH10_LaBH8_CP}b -- but stable MH$_{8}$ phases with this symmetry have recently been predicted for several of the smaller rare earth elements including Nd, Ho, Er, and Tm~\cite{Sun:2020}. 

\begin{figure*}
\begin{center}
\includegraphics[width=1.3\figurewidth]{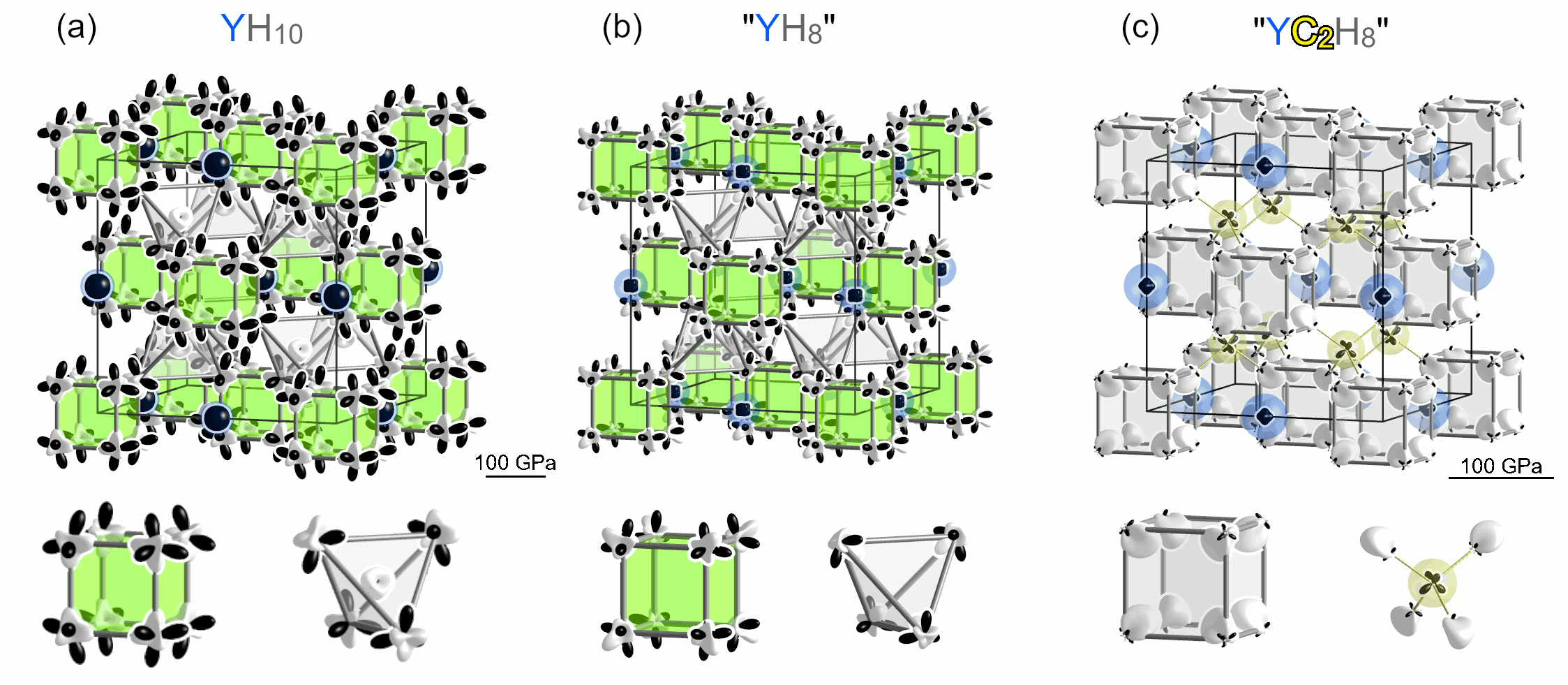}
\end{center}
\caption{Chemical pressure schemes of yttrium hydrides with $Fm\bar{3}m$ symmetry based on a hydrogenic lattice of vertex-sharing tetrahedra and cubes (insets). In YH$_{10}$ (a) the tetrahedra are each stuffed with an additional H atom. The Y atoms display negative pressure, combined with strong positive H-H pressures that indicate a too-large coordination environment for the Y, which forces the hydrogenic lattice into an overly-contracted configuration. Removing the H atoms from the tetrahedra yields the model ``YH$_{8}$'' structure (b) that retains the negative pressure on Y and positive pressures between H atoms. A proposed structure, YC$_{2}$H$_{8}$ (c) has carbon atoms stuffed into the tetrahedra and leaves the cubes empty.
\label{fig:YHx_CP}}
\end{figure*}

Despite numerous attempts $Fm\bar{3}m$ YH$_{10}$, predicted to be superconducting at room temperature~\cite{Peng:2017a}, has yet to be synthesized. Could the CP method provide clues on the elusiveness of this phase, and suggest appropriate doping strategies to stabilize it to low pressures? Because of its smaller radius, the CP map of $Fm\bar{3}m$ YH$_{10}$ (Figure ~\ref{fig:YHx_CP}a) differs markedly from that of LaH$_{10}$. Notably, the Y atom is surrounded by negative CP, indicating that it is too small for its coordination environment, whereas most of the H-H contacts are too short, displaying positive CPs. The most compressed H-H contacts are those comprising the H$_{8}$ cubes, as well as those between the H atoms forming the tetrahedra and those stuffed within them. Negative pressure features on the H atoms are solely directed towards the Y atoms, thus showing that the overall tension lies between a cramped H cage being pulled in by a too-small Y atom at its center. The different stresses inherent in the La- and Y-based MH$_{10}$ hydrides suggest a mixture of La and Y may stabilize the framework, as demonstrated by a ternary (La,Y)H$_{10}$ phase, which was detected alongside the substituted (La,Y)H$_{6}$ $Im\bar{3}m$ structure~\cite{Semenok:2021}.

As with the La phases, we can remove the H2 atoms at the center of the tetrahedra to create an $Fm\bar{3}m$ ``YH$_{8}$" model structure (which differs from the predicted metastable $Cc$ YH$_8$ phase~\cite{Liu:2017}) whose CP map may reveal where to place additional atoms to stabilize the Y-H framework. Figure \ref{fig:YHx_CP}(b) shows that the Y atoms remain surrounded by negative pressures, mismatched with an H coordination environment that is still too large. Among the H atoms, positive pressures lie along the H-H contacts, with the largest being directed towards the center of the tetrahedra. This leads to an interesting situation -- expanding the size of the H$_{8}$ cubes through the addition of a stuffing element such as B may only contract the tetrahedra even further, exacerbating the positive pressures already present in that space. Moreover, if, analogous to what was found for the lanthanum analogue, stuffing a B atom into YH$_{8}$ leads to an even larger negative pressure around the metal atom as compared to within YH$_{10}$, this would further destabilize the structure. In agreement with these qualitative arguments, we found that a YBH$_8$ phase isotypic with LaBH$_8$ was not dynamically stable in the pressure range of 0-200~GPa.

Since carbon has a radius that is smaller than boron we postulated that it could be used to stabilize the hydrogenic network instead. In the resulting $Fm\bar{3}m$ TC$_{2}$H$_{8}$ phase (Figure \ref{fig:YHx_CP}(c)) the carbon atoms stuff the tetrahedral vacancies, resulting in methane molecules with strong C-H bonds, as confirmed by a plot of the ELF (Fig.\ S20), spread throughout an fcc yttrium lattice. One important consequence of the presence of the C-H bond can be seen by comparing the H-H distances along the  H$_{4}$  tetrahedral edges. 
Because hydrogen atoms fill the tetrahedra in YH$_{10}$, this contact measures 2.11~\AA{}, shrinking to 2.00~\AA{} after their removal in YH$_8$. Upon insertion of carbon into the tetrahedra the H-H distance \emph{decreases} further to 1.88~\AA{}. Typically, it would be expected that stuffing an atom into a cluster would cause it to expand, but in this case it shrinks due to the formation of covalent C-H bonds within the methane molecule. The shrinking of the tetrahedra relieves some of the tension in the H$_{8}$ cubes with which they share vertices, thereby stabilizing the phase. Our results highlight that the choice of the appropriate dopant atom required to stabilize a particular environment depends both upon steric factors, and the potential for the formation of covalent bonds that might shorten selected contacts. In the resulting CP scheme of YC$_{2}$H$_{8}$, the C-H contacts within the methane-like fragments are nearly optimized, with the C mainly displaying negative pressure towards the Y atoms. Negative CP lobes also decorate the Y-H contacts, reflecting Y encapsulation in a slightly-too-large environment, also seen in the YH$_{10}$ and ``YH$_{8}$'' schemes. A similar tactic of heteroatom substitution into the tetrahedral spaces in the hydrogen lattice is employed in the proposed LaC$_2$H$_8$~\cite{Durajski:2021} and KB$_2$H$_8$~\cite{Gao:2021} phases, whose $T_c$s are predicted to be 69~K at 70~GPa and 146~K at 12~GPa, respectively.

YC$_{2}$H$_{8}$ is dynamically stable to pressures as low as 50~GPa (Fig.\ S21), and it is metallic, with an appreciable density of states at the Fermi level, suggesting it may be superconducting. Calculations revealed a sizeable electron phonon coupling parameter, $\lambda$=3.3, in which a majority of the contributions were from the three acoustic phonon modes associated primarily with motions of the Y atoms, and CH$_4$ fragments moving mainly as single units, with a particularly sizeable contribution along the K-$\Gamma$ path (Fig.\ S21). Combined with an average logarithmic frequency of $\omega_\text{log}$=132~K, this yields a $T_c$ of 25~K (24~K) as estimated by the Allen-Dynes modified McMillan equation using a Coulomb pseudopotential, $\mu^*$, with a typical value of 0.1 (0.13). Due to the especially strong electron-phonon coupling, the $T_c$ was also estimated via numerical solution of the Eliashberg equations, yielding values of 61~K (54~K) for $\mu^*$=0.1 (0.13).

To balance the simultaneous requirement of high $T_c$ coupled with low pressure required for a technologically useful material, a superconducting figure of merit, $S$, has been proposed~\cite{Pickard:2020}. In this scale the highest $T_c$ conventional superconductor at 1~atm that is known,  MgB$_2$, possesses $S$(MgB$_2$)~=~1, and a perfect value of 10 would be associated with a material whose $T_c$~=~390~K at ambient pressure. On this scale $S$(YC$_2$H$_8$)~=~0.96, similar to MgB$_2$ and reflective of the moderate increase in $T_c$ being matched by a moderate increase in pressure. In comparison $S$(YH$_{10}$)~=~1.29, and no ternary hydrides derived from $Fm\bar{3}m$ YH$_{10}$ with a higher $S$ have been proposed to date.
Assessing the thermodynamic stability of YC$_{2}$H$_{8}$ as a function of pressure and temperature would require first-principles based crystal structure prediction searches in the Y-C-H ternary phase diagram, followed by free energy calculations on the most promising phases. This computationally demanding protocol is past the scope of the current study, and it is currently not possible to carry out such a procedure for every potential ternary combination. Instead, we propose that the CP-based technique illustrated here can first be employed to rationally design ternary compounds that are likely to superconduct at high temperatures followed by more extensive explorations of potential synthetic pathways.

\section*{Conclusion} 

Ever since the superhydride superconductors have been predicted, and some have been synthesized, many questions, whose answers would facilitate the discovery of more complex hydride based superconductors, remained unanswered. Perhaps the most pressing are: ``Why do certain metal atoms favor the formation of the hydrogenic clathrate cages key for the high superconducting critical temperatures ($T_c$s)?'', and ``Can we rationally alter the superhydrides to design high-$T_c$ materials stable at low pressures?'' Herein, the DFT-Chemical Pressure (CP) method is used to answer both of these questions. First, we showed how the size of the electropositive metal atom is linked to the structure the hydrogen lattice adopts, focusing on the ubiquitous MH$_6$ stoichiometry. Our detailed analysis explains why the high-symmetry $Im\bar{3}m$ structure with record-breaking $T_c$ values is stable when the hydrogen clathrate cage encapsulates Mg, Ca or Sc atoms, but not Sr, Ba or La. Next, we demonstrated how the DFT-CP method can be used to identify void spaces within the hydrogenic framework where the addition of a stuffing atom decreases the internal strains within the structure. Now that near room temperature superconductivity has been measured, the immediate goal  is to find materials that can maintain high $T_c$ to ambient pressures~\cite{Zurek:2020k}. Because the prospect of thoroughly searching the phase diagram of every possible ternary composition is untenable, techniques to rationally design promising systems, like the CP method, are urgently required. The LaH$_{10}$/LaBH$_{8}$ and YH$_{10}$/YC$_{2}$H$_{8}$ families are two examples, in which a third element -- B, C, or additional H -- serves as a stuffing atom in void spaces that can be identified by the CP method, left by the hydrogenic network. The proposed YC$_{2}$H$_{8}$ phase represents an example of this CP-driven materials design for low pressure, high temperature metal hydride superconductors, which will be applied to uncover further exemplary materials in future studies.

\section*{Acknowledgments}

K.H. is thankful to the U.S.\ Department of Energy, National Nuclear Security Administration, through the Capital-DOE Alliance Center under Cooperative Agreement DE-NA0003975 for financial support.  We acknowledge the NSF (DMR-1827815) for financial support. Calculations were performed at the Center for Computational Research at SUNY Buffalo \\ (http://hdl.handle.net/10477/79221).

\bibliography{ClathrateCP}

\clearpage
\newpage

\end{document}